# Emergent self-propulsion at low Reynolds number

Alexander Reinmüller,*[a] Hans Joachim Schöpe[a] and Thomas Palberg[a]

[a] *Institut für Physik, Johannes Gutenberg-Universität Mainz,*
*Staudingerweg 7, 55128 Mainz, Germany*

* Electronic mail: *reinmu01@uni-mainz.de*

We here demonstrate the experimental realization of inanimate micro-swimmer complexes showing emergent self-propulsion at low Reynolds number (Re < $10^{-4}$) in quasi 2D colloidal systems. Guided by the substrate, self-organized propulsion occurs for up to 25min with typical velocities of 1-3μm/s, while none of the involved constituents shows self-propulsion on its own.

Self-propulsion at low Reynolds numbers is of principal importance for micro-organisms and engineering purposes as well as fascinating from the fundamental point of view [1]. Exploiting various physico-chemical mechanisms, individual or mechanically coupled autonomous micro-swimmers in liquid media show locomotion outbalancing diffusive motion in terms of velocity and directedness [2-5]. In the present paper, we demonstrate that self-propulsion may also emerge from co-operative dynamics in a self-assembled multi-constituent complex [6]. Such complexes comprise of micron-sized colloidal spheres accumulated at larger, electrolyte releasing particles on a charged substrate. Neither isolated constituent shows motility beyond slow Brownian motion. Self-assembly is intrinsically driven by an electrolyte-gradient-induced, annular electro-osmotic solvent flow. The convection cell is essential for self-propulsion of the assembled complex in the substrate surface plane. The onset of motion is triggered by a colloid density asymmetry. Once launched, propulsion is self-stabilizing. The mechanism facilitates collecting and transporting colloidal objects loosely bound within the asymmetrically loaded convection cell over considerable distances.

In our experiments, we investigated aqueous mixtures of colloidal microspheres and cation exchange resin (CIEX) fragments settled onto a glass substrate using optical microscopy. Sample preparation was performed at room temperature using commercially purchased, negatively charged Polystyrene spheres of diameter $2a = 5.2$μm (Batch No. PS/Q-FB1036 by MicroParticles Berlin GmbH, Germany) in milliQ grade water. Cation exchange resin (CIEX) spheres (Amberlite K306, Roth GmbH, Germany) were crushed and a few tiny fragments (≤ 100μm) were placed on a cleaned glass microscopy slide. Very dilute colloidal suspensions (volume fraction < 0.1%) were added. Drops were in contact with room air providing a residual electrolyte concentration of approx. 5μmol/l. In some cases the drop was covered with another glass slide using spacers, but in other cases no upper substrate was used. The fluid heights in both cases amounted to several 100μm. Both colloidal particles and CIEX fragments were subject to gravity and settled within minutes. For optical microscopy a standard scientific microscope (DM-IRBE by Leica, Germany) was used equipped with low magnification (10× and 20×) objectives and a standard color CMOS camera (SMX-M73 by EHD imaging GmbH, Germany). Image sequences were the basis for further quantitative analysis (Supplementary Note: Correlations).

First, we observed the spontaneous accumulation of the colloidal particles around isolated CIEX fragments. Radially symmetric attractive forces towards the CIEX fragments acting on settled colloidal particles were observed in a range of several tens to hundreds of microns. [7] The attraction strength and range generally increase with the CIEX fragment size, which leads to a rich variety of differently appearing particle assemblies (Fig. 1).

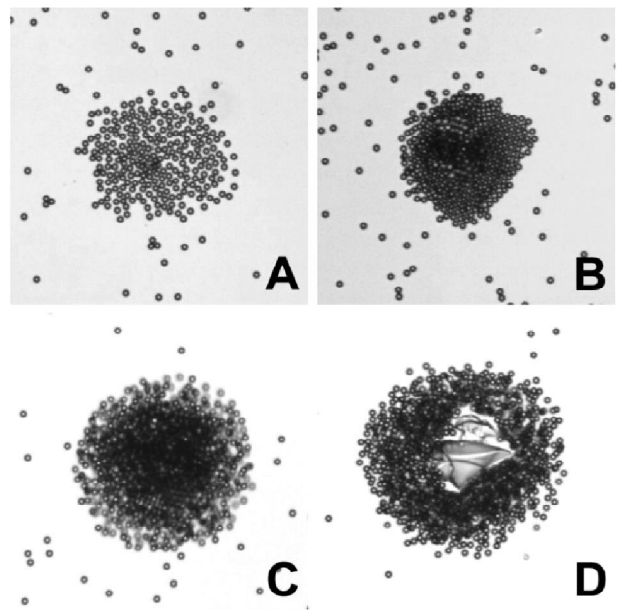

**Fig. 1** Optical micrographs showing typical colloidal particle accumulations around stationary cation exchange resin (CIEX) fragments on a glass substrate: Loosely crowded monolayer (A), crystalline double layer (B), and marked convective circulation (C, D). A huge CIEX fragment is clearly visible in (D). (Scale: 265×265μm²)

In the case of 'weaker' CIEX fragments (typically < 10µm), usually disordered (Fig. 1A) or crystalline ordered (Fig. 1B) particle assemblies of one or two sedimented layers emerged as a result of local lateral particle accumulation. In the case of 'stronger' CIEX fragments (typically > 10µm), the colloidal particles followed significant convective solvent currents (Figs. 1C, 1D): The particles first laterally converged in the substrate surface plane. At the CIEX, they then followed the upwards directed flow to altitudes of up to ca. 30µm above the substrate surface against gravity. After several tens of seconds they fell, laterally displaced, back to the substrate, where they again were swept inwards by the convective solvent flow. Under sufficiently large particle influx, prominent particle circulation evolved.

Many of those complexes of colloidal particles around CIEX fragments were stationary. But especially in some cases of larger CIEX, which usually exhibited pronounced convection, directed lateral motion was observed. This motion was correlated to an asymmetric particle distribution around the CIEX fragment. An exemplary snapshot is given in Fig. 2: An elongated, irregularly shaped CIEX fragment is loosely, but asymmetrically surrounded by settled colloidal particles. These approach the CIEX from all directions. Nearby the CIEX, they leave the focal plane following the vertical convective flow upwards. Immediately behind the CIEX, particles have accumulated within a convective circulation cell, while farther behind a densely crowded trail of settled particles has formed. The whole complex moves linearly with an average velocity of v = (1.3±0.1) µm/s.

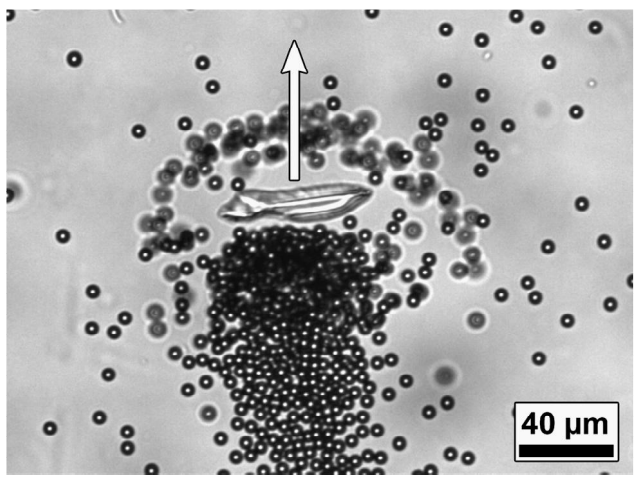

**Fig. 2** Optical micrograph of a self-propelling complex: It consists of colloidal particles (circular objects) accumulated at a CIEX fragment (bright elongated object in the image center); the direction of motion is indicated by the arrow. Colloidal particles above the sedimented layer appear blurred. (Scale: 280×210µm²)

Propulsion sets in, when markedly asymmetric particle distributions are formed due to fluctuations out of a symmetric initial state without any significantly directed motion. The majority of the colloidal particles always followed behind the moving CIEX fragment. This configuration is self-stabilizing, since *oncoming* particles fall down *behind* the CIEX, after they have been lifted up by the convective flow. Some of these particles remain bound in the back circulation cell, but excess particles loose contact and form a characteristic trail. Besides nearly linear trajectories (Fig. 3A), also curved (Figs. 3B, 3C) and even circular trajectories (Fig. 3C) were observed, which slowly dissolved by diffusion.

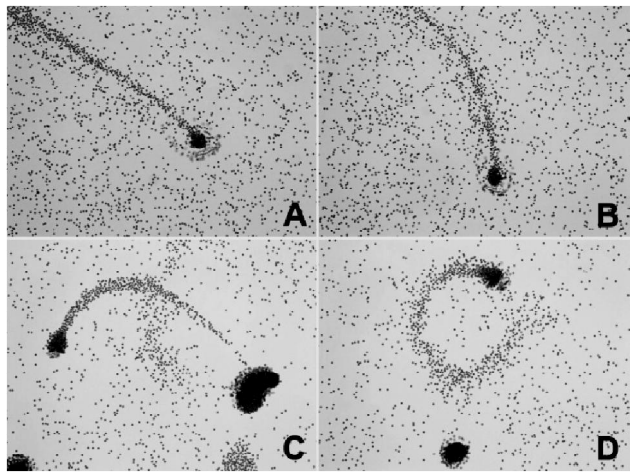

**Fig. 3** Optical micrographs of micro-swimmers with differently curved trajectories. (A) and (B) show the same swimmer at different times. (Scale: 1150×860µm²)

At moderate particle influx from the surroundings, steady state behavior was observed over several minutes. Within these, the swimmers covered path lengths of many hundreds of microns. Typical velocities were between 1 and 3µm/s which significantly outbalance the diffusive motion of a single colloidal particle (Stokes-Einstein diffusion coefficient ~ 0.09µm²/s).

On timescales of 10min as well as under marked particle influx, a characteristic temporal evolution of the system dynamics became observable: After launching propulsive motion, an acceleration period follows, which features a particle number increase within the back convective circulation together with a negligible loss of particles lagging behind. Then a period of maximum speed follows, which usually exhibits the characteristic trail. Motion either stops abruptly, when the CIEX fragment is adhered to an impurity on the substrate. Then the particle distribution quickly relaxes to a rather symmetric state, while convection still goes on (Fig. 1 C and D). Or the swimmer decelerates continuously, seemingly under particle overload.

Preliminary evaluation of self-propulsion events already reveals some rough trends in the dependencies on experimental parameters: Moving

CIEX fragments always had irregular shapes, some more elongated, but others more rounded. They always moved long side ahead. Larger velocities were weakly correlated with larger CIEX fragments, more pronounced and extended particle circulation and larger particle number densities in the surrounding fluid. But in principal, under too large particle number densities in the substrate surface plane, no propulsive motion was observed. In this case, only stationary, rather symmetric particle accumulations were found. In the case of too small particle number densities, no significant particle accumulation could evolve. Then the CIEX fragments performed only small scale irregular motion.

This self-propulsion effect is powered by intrinsically generated electro-osmosis in the surface plane of the charged glass substrate. In a recent publication [7], we could demonstrate the release of electrolyte by isolated ion exchange resin spheres. In particular, the CIEX releases the residual activator substance HCl - hydrochloric acid. This induces significant pH-changes by 2-3 orders of magnitude in the local surrounding over a range of several 100µm within a few minutes. Dissociation and faster diffusion of the $H^+$ ions with respect to $Cl^-$ lead to the formation of concentration gradients. Direct diffusio-phoretic interaction of these electrolyte gradients with the colloidal particles [8] could be shown to be of minor importance here [7]. Rather, the associated gradients in osmotic pressure, charge density and pH interact with the counter-ions of the charged glass substrate to induce an electro-osmotic flow along the substrate. [3, 8] In our geometry, this flow radially converges at the CIEX fragment. Solvent incompressibility further enforces an upward flow above the CIEX fragment, which can nicely be traced using colloidal particles in combination with large CIEX fragments. Significant changes in the electrolyte release by the CIEX, possibly indicated by gradual, irreversible adhesion of colloidal particles on the CIEX, may weaken the convective pattern and slow the complex.

Our observations show that, basically, the annular convective flow is radially symmetric. Since no propulsion was observed under symmetric particle distributions or under absence of particles, we believe that the lateral force balance is broken by asymmetric particle distributions. This assumption is corroborated by the observation of collisions like that in Fig. 4, where a change in direction of motion away from large particle densities has happened. Hence, the local particle distributions may determine both speed and direction of motion. We further suggest that self-propulsion happens regardless of the state of motion of the accumulated objects. This suggestion is based on the observation of one micro-swimmer complex which dragged a tail of slowly moving agglomerates behind. Those consisted of both colloidal particles and small CIEX fragments and did not participate in the solvent circulation.

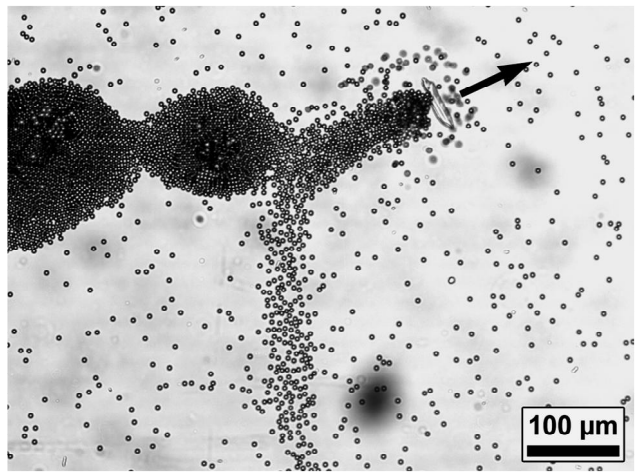

Fig. 4 Optical micrograph of a micro-swimmer (the same from Fig. 2, but two minutes later) after a collision with a stationary, dense particle assembly: The direction of motion (arrow) has changed. (Scale: 720×540µm2)

At least one possible mechanism may explain the observed propulsion quantitatively: Motion may be driven by an osmotic pressure imbalance associated with the particle accumulation. Assuming the osmotic pressure to be equivalent to an excess hydrostatic pressure, the particle concentration gradient exerts a pressure gradient force on the CIEX fragment. Applying an analytical model together with physically reasonable parameters we estimate micro-swimmer velocities in the range $v \sim 1\text{-}10\mu m/s$ (Supplementary Note: Velocity estimate from the osmotic pressure mechanism) which coincides well with the experimental observations. However, due to uncertainty about intrinsic experimental conditions as well as involved boundary conditions, further mechanisms may be conceivable. E. g., presence of particles may unilaterally alter the hydrodynamic and rheological properties of the fluid, or due to their charge, they may unilaterally influence strength and distribution of the electrolyte gradient, what may both lead to a net force. Extrinsic forces resulting from macroscopic external fields, gradients or capillary forces, however, could safely be ruled out, since motion of different micro-swimmers in the same system were basically not correlated with each other, and since the colloidal particles around did not show any motion except thermal fluctuations.

In conclusion, we here presented an innovative microscopic particle collection and transport mechanism under low Reynolds number conditions exploiting intrinsic, locally generated electro-osmotic currents. In contrast to an individual micro-swimmer particle, the basic requirements for autonomous self-propulsion are here implemented in different co-operative components that do not show self-propulsion on their own. First, the *motor* needed to generate the driving force is provided by electro-osmosis in the surface plane of the charged glass substrate. Second, the *reservoir* needed to store and release fuel substances in sufficient amounts is provided by the CIEX resin particle. Third, the *gearing and steering*, determining direction and velocity of the swimmer are realized through the accumulated colloidal particles, while gravity holds the complex close to the substrate. This restricts the complex´ motion to quasi 2D. This modular structuring may provide particular flexibility in designing each component individually for an optimum contribution to the observed collective self-organized propulsion as well as for specific transport applications. These first experiments were comparably simple and did not require elaborate system preparation which indicates a rather robust mechanism. This kind of emergent propulsion seems to require only a sufficiently strong annular convection cell about a sufficiently mobile, fuelled particle, which is asymmetrically loaded by cargo particles. However, the underlying dynamics and particularly the emergent behavior are not yet understood in a quantitative way. Our observations therefore also pose challenging questions for both further experimental studies and theoretical modeling.

## Acknowledgements

We gratefully acknowledge financial support by SFB-TR6 projects D1 and B1 by Deutsche Forschungsgemeinschaft.

## Supplementary Note: Correlations

The following quantities of the experimentally observed micro-swimmers were estimated using conventional image analysis tools: The velocity $v = \Delta s/\Delta t$ was estimated using the linear CIEX displacement $\Delta s$ in a time interval of typically $\Delta t = 20s$. The CIEX size in terms of its projected area $A_{CIEX}$ in the frame was estimated using the freehand selection tool of common open source image analysis software (Image J, currently available at 'http://rsbweb.nih.gov/ij/'). Likewise, the projected area $A_{conv}$ of the back convection cell including densely crowded colloidal particles was estimated as a measure for the convective flow strength. The particle area number density $n_A$ in the surrounding dilute suspension (neglecting the densely crowded trail) was estimated using a particle counting software tool (Image J). The CIEX aspect ratio was further calculated by dividing the measured CIEX width by its length with respect to the flow direction. Long sequences were segmented into shorter sequences which were evaluated independently in order to allow for temporal evolution of the micro-swimmers. No stationary or slowly fluctuating CIEX ($v < 0.5$ µm/s) were evaluated.

For geometrical parameters, we obtained the following rough dependencies: Larger velocities $v$ are weakly correlated with larger CIEX sizes $A_{CIEX}$ (Supplementary Figure 1A). Further, larger velocities are induced at larger $n_A$ (Supplementary Figure 1B) as well as at more extended convection cells $A_{conv}$ (Supplementary Figure 1C). No evident correlation between the velocity and the CIEX aspect ratio is observed (Supplementary Figure 1D). Increasing convection cell areas $A_{conv}$ on the one hand correlate with increasing CIEX sizes $A_{CIEX}$ (Supplementary Figure 2A) but on the other hand with decreasing aspect ratios (Supplementary Figure 2B). The complex dependency between $A_{conv}$ and $n_A$ (Supplementary Figure 2C) indicates a non-trivial interplay of experimental parameters. The data points at $A_{conv} = 0$ correspond to micro-swimmer complexes that did not show obvious particle circulation. The data point at $n_A = 0$ corresponds to a single observation of a micro-swimmer that passed a small particle-free area in a very inhomogeneous particle distribution. The evaluated parameters may not fully characterize a micro-swimmer, since *local* chemical parameters like the ion concentration profile around the CIEX or its ion release rate could not be determined.

**Supplementary Note: Velocity estimate from the osmotic pressure mechanism**

Based on an analytical model [9], we quantitatively estimate the osmotic pressure difference using physically reasonable parameters. The osmotic pressure $P$ of low salt charge-stabilized colloidal suspensions is mainly determined by the counter ions. For non-vanishing salt concentrations $c_s$, integrating

Eq. 50 from reference [9] yields $P = A 2\pi\lambda_B \rho_c^2 Z^{*2}/(\kappa^2 \beta)$. The particle number density in the accumulated regions is assumed to be of the order of magnitude $\rho_c \sim 10^{15}/m^3$ which corresponds to volume fractions $\varphi \sim 10\%$. For the system used here, the effective particle charge is of the order of magnitude $Z^* \sim 10^5$ as obtained by conductivity measurements [10]. The Bjerrum length is denoted by $\lambda_B$, and $\beta$ denotes the inverse thermal energy. With the squared screening parameter, $\kappa^2 = 4\pi\lambda_B(2N_A c_s + Z^* \rho_c)$, the background salt concentrations between $c_s \sim 5\mu$mol/l ($CO_2$ saturation without HCl) and $c_s \sim 10^{-4}$mol/l (pH = 4 due to HCl release) correspond to screening parameters (inverse screening lengths) between $\kappa \sim 7/\mu$m and $\kappa \sim 30/\mu$m. In this range, the pre-factor, $A = 1+(\kappa a)^2/(3(1+\kappa a)^2)$ amounts to $A \sim 1.3$. Using these values, we obtain osmotic pressures between $P \sim 10^{-3}$Pa and $P \sim 10^{-4}$Pa which may act on the CIEX fragments equivalently to an excess hydrostatic pressure.

We further consider a spherical CIEX fragment of radius $a_c = 30\mu$m which is subject to Stokes drag and owns a convection cell loaded on one side only. With the viscosity $\eta = 10^{-3}$Pas of water, the pressure gradient force $F = P \times \pi a_c^2 = 6\pi\eta a_c v$ acting on the sphere leads to micro-swimmer velocities between $v \sim 10\mu$m/s and $v \sim 1\mu$m/s, which coincide with the experimental values.

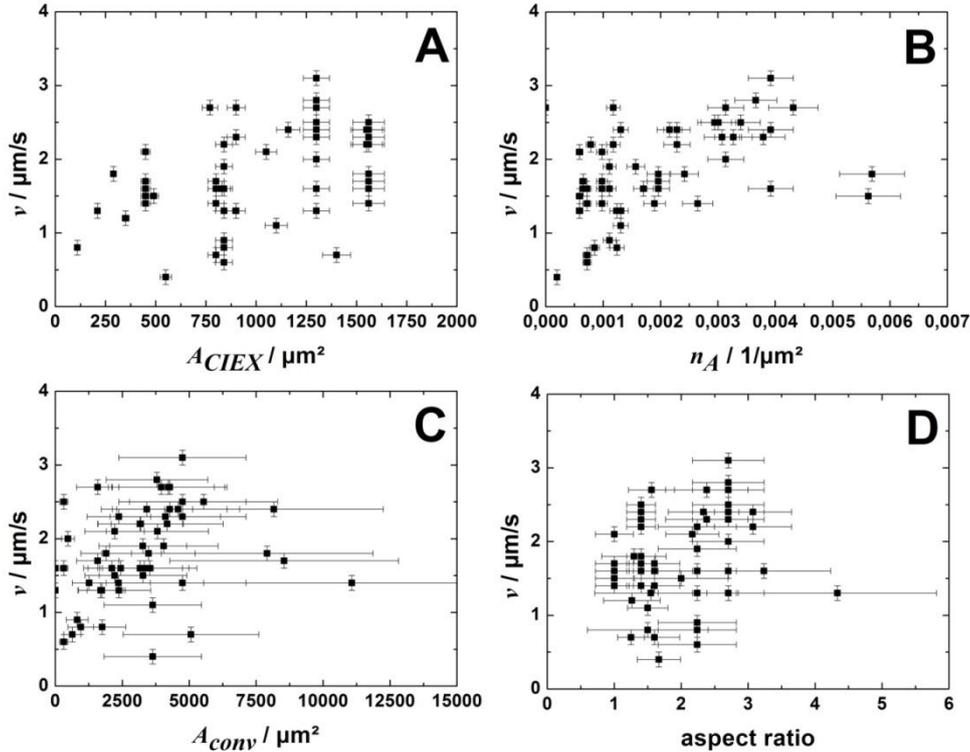

**Supplementary Figure 1.** Correlations between the micro-swimmer velocity and the CIEX size (A), the surrounding area number density (B), the convection cell area (C) and the CIEX aspect ratio (D).

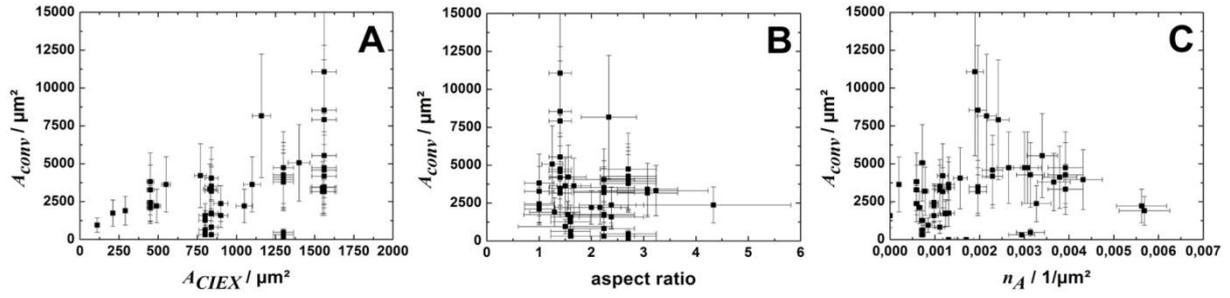

**Supplementary Figure 2.** Further correlations between the convection cell area and the CIEX size (A), the CIEX aspect ratio (B) and the surrounding area number density (C).